\documentclass[aip,ajp,preprint,longbibliography]{revtex4-1}
\usepackage{amssymb}
\usepackage{amsmath}
\usepackage[colorlinks]{hyperref}
\usepackage[normalem]{ulem}
\usepackage{graphicx}
\usepackage{mathrsfs}
\begin{document}

\title{An algebra and trigonometry--based proof of Kepler's First Law}
\author{Akarsh Simha}
\affiliation{Sunnyvale, CA, USA}
\email{akarsh@utexas.edu}
\begin{center}\emph{Published in the American Journal of Physics \textbf{89}, 1009 (2021)}\end{center}

% More TikZ stuff
% \contourlength{1.2pt}
% %% Copied from: https://tex.stackexchange.com/questions/20826/label-angle-with-tikz
% \newcommand{\tikzAngleOfLine}{\tikz@AngleOfLine}
%   \def\tikz@AngleOfLine(#1)(#2)#3{%
%   \pgfmathanglebetweenpoints{%
%     \pgfpointanchor{#1}{center}}{%
%     \pgfpointanchor{#2}{center}}
%   \pgfmathsetmacro{#3}{\pgfmathresult}%
% }
\newcommand{\point}[1]{ \fill[gray] {#1} circle (1pt); }
\newcommand{\strongpoint}[1]{ \fill[black] {#1} circle (1pt); }

\begin{abstract}
  An elementary proof of Kepler's first law, i.e. that bounded planetary orbits are elliptical, is derived without the use of calculus. The proof is similar in spirit to previous derivations, in that conservation laws are used to obtain an expression for the planetary orbit, which is then compared against an equation for the ellipse. However, we derive the equation that we match against, using trigonometry, from two well-known properties of the ellipse. Calculus is avoided altogether.
  % makes use of conservation laws to obtain an unusual equation of the ellipse, but calculus is avoided altogether. A trigonometric derivation of the equation describing the ellipse from simple properties is provided, along with visual tools, making this presentation self-contained.
  % ~\cite{vogt1996elementary,noll2002teaching}
\end{abstract}

\maketitle

\section{Introduction}
\label{sec:Introduction}
The explanation of planetary orbits by Newton was one of the major triumphs of his laws of motion and of gravitation. Elementary physics textbooks at the high school and undergraduate college level invariably describe Kepler's three laws of planetary motion in their chapters on gravitation, but most textbooks only derive the result for circular orbits. Proofs that avoid calculus are a valuable teaching aid, appealing to both students of physics and keen amateur physicists and astronomers. Several elementary derivations have been provided by others, based on calculus, geometry, or algebra and trigonometry.~\cite{vogt1996elementary,noll2002teaching,goodstein1996feynman,provost2009simple,unruh2018kepler} However, analytical derivations seen in the literature tend to reduce the dynamical equations to an equation of the ellipse that is not typically encountered by the student or enthusiast in their coursework, and the justification that this equation describes an ellipse has either been taken as a standard mathematical result,~\cite{noll2002teaching} or has been derived by applying calculus.~\cite{vogt1996elementary}

While it is still true in this work that we employ an unusual equation of the ellipse, we use trigonometry to derive this equation from two well-known properties. This equation is then readily matched against an equation derived by applying conservation laws to the physical problem, showing that the orbit is elliptical. We believe, therefore, that the assumptions from which we derive the results in this work should be relatively easy for a physics undergraduate or enthusiast to accept.

\section{Conservation of Energy and Angular Momentum}
\label{sec:ConservationLaws}
Conservation laws play an important role in describing the dynamics of systems by providing us with constants of motion. In fact, if a sufficient number of independent constants of motion can be determined, the dynamics of a system can be completely described.\footnote{The Liouville-Arnold integrability theorem for Hamiltonian systems relies on finding as many independent constants of motion in \emph{involution} as there are degrees of freedom in the system.} There are three conserved quantities in the Kepler problem, but we shall concern ourselves with only two of them here, energy and angular momentum.

\begin{figure}
  \begin{center}
  \includegraphics{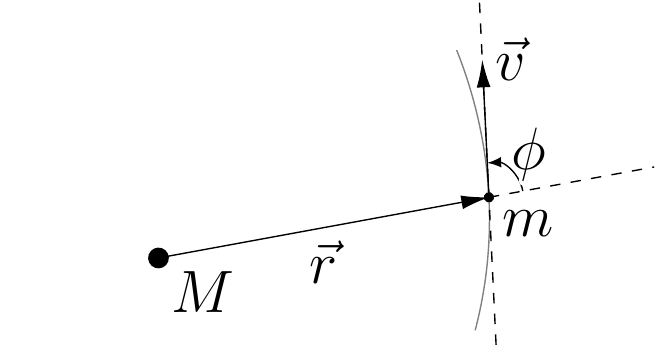}
\caption{A figure representing the physical system under consideration. A mass $m$ orbits another mass $M \gg m$. The origin is chosen to coincide with the position of $M$, which is approximated to be stationary. The velocity of $m$ is denoted by $\vec{v}$ and its position by $\vec{r}$. The vector $\vec{v}$, which is tangent to the trajectory of $m$, makes an angle $\phi$ with the vector $\vec{r}$.}
\label{fig:PhysicalSetup}
  \end{center}
\end{figure}

To simplify the Kepler problem, we shall assume that one body (``the Sun'') has a mass $M$ that is \emph{much} larger than the mass $m$ of the other body (``the planet''), so that $M$ may be treated as if it were stationary in space.\footnote{A formulation using relative coordinates and reduced mass may be followed to overcome this approximation if needed} Placing $M$ at the origin, let us denote the position of the moving body $m$ by vector $\vec{r}$, and its velocity by $\vec{v}$. Let $\phi$ be the counterclockwise angle from $\vec{r}$ towards $\vec{v}$. This setup is illustrated in Fig.~\ref{fig:PhysicalSetup}. We may then write the angular momentum of the body as
\begin{equation}
  \label{eq:AngularMomentum}
  \vec{L} = m \vec{r} \times \vec{v} = m r v \sin \phi \hat{z},
\end{equation}
where $r$ and $v$ denote the magnitudes of $\vec{r}$ and $\vec{v}$, respectively, and $\hat{z}$ is a unit vector in the direction perpendicular to $\vec{r}$ and $\vec{v}$. Since the gravitational force acting on the body $m$ acts along the line joining the bodies, no torque is exerted on the body, whereby the angular momentum $\vec{L}$ is conserved, both in magnitude $L$ and direction $\hat{z}$. The constancy of the direction tells us the trajectory must lie in a plane perpendicular to $\hat{z}$.

Since there are no non-conservative forces acting on the body $m$, the energy $E$ of the system is also a constant of motion. It is given by the sum of kinetic and gravitational potential energies,
\begin{equation}
  \label{eq:Energy}
  E = \frac{1}{2} m v^2 - \frac{G M m}{r},
\end{equation}
where $G$ is the universal gravitational constant. Note that we have followed the convention where gravitational potential energy is $0$ at infinite separation, whereby the total energy $E$ of a bound orbit will be negative (since such a body does not have enough energy to reach infinity). We are confining our treatment here to the case of bounded orbits.

\section{The equation of trajectory from conservation laws}
\label{sec:Trajectory}
Solving Eq.~\eqref{eq:Energy} for $v$, we obtain
\begin{equation}
  \label{eq:Velocity}
  v = \sqrt{\frac{2E}{m} + \frac{2GM}{r}}.
\end{equation}
Plugging this into~\eqref{eq:AngularMomentum}, we obtain
\begin{equation}
  \label{eq:PreTrajectory}
  L = m  \sin \phi \sqrt{\frac{2Er^2}{m} + 2 G M r},
\end{equation}
which may be rearranged as
\begin{equation}
  \label{eq:Trajectory}
  \left ( r^2 + \frac{GMm}{E} \, r \right ) \sin^2 \phi = \frac{L^2}{2mE}.
\end{equation}

The above equation tells us how the angle $\phi$ between the position and velocity vectors of $m$ varies with the distance $r$. Noting that the velocity vector is tangent to the trajectory, we identify $\phi$ with the angle made by the tangent to the position vector of the planet. Thus, in principle, Eq.~\eqref{eq:Trajectory} describes the orbit traced out by the planet. However, it is presented in terms of unusual variables, and hence cannot be matched immediately to a known standard equation.

\section{An equation describing an ellipse in terms of its tangent}
\label{sec:PedalEquation}

In this section, we shall derive an unusual equation of an ellipse that will immediately lead to Kepler's first law when matched against Eq.~\eqref{eq:Trajectory}. Let us denote the \emph{semi-major axis} of the ellipse by $a$ and its \emph{eccentricity} by $e$. One definition of the eccentricity $e$ is that the foci lie at a distance of $ae$ on opposite sides of the center of the ellipse. Since the foci lie inside the ellipse, $0 \leq e < 1$, the case $e = 0$ of coincident foci corresponding to a circle.

%As is well known, an ellipse is a closed plane curve informally thought of as a stretched-out circle with different extents along two perpendicular dimensions. The quantity $a$ is known as the semi-major axis, and $b$ is the semi-minor axis. Informally, $2a$ is the length of the longer side and $2b$ is the length of the shorter
% side of a rectangle that tightly bounds the ellipse. The ellipse has two special points called the \emph{foci}, denoted in the figure~\ref{fig:PedalEquation} as $F$ and $F'$. These are at a distance $ae$ away from the center of the ellipse on either side along the major axis, where the quantity $e$, which is always less than 1 for an ellipse, is known as the \emph{eccentricity}. 
% The quantities $a$, $b$ and $e$ are related through:

% The semi-minor axis $b$ is related to $a$ and $e$ through the expression:
% \begin{equation}
%   \label{eq:Eccentricity}
%   b^2 = a^2 (1 - e^2).
% \end{equation}

\begin{figure}
  \begin{center}
  \includegraphics{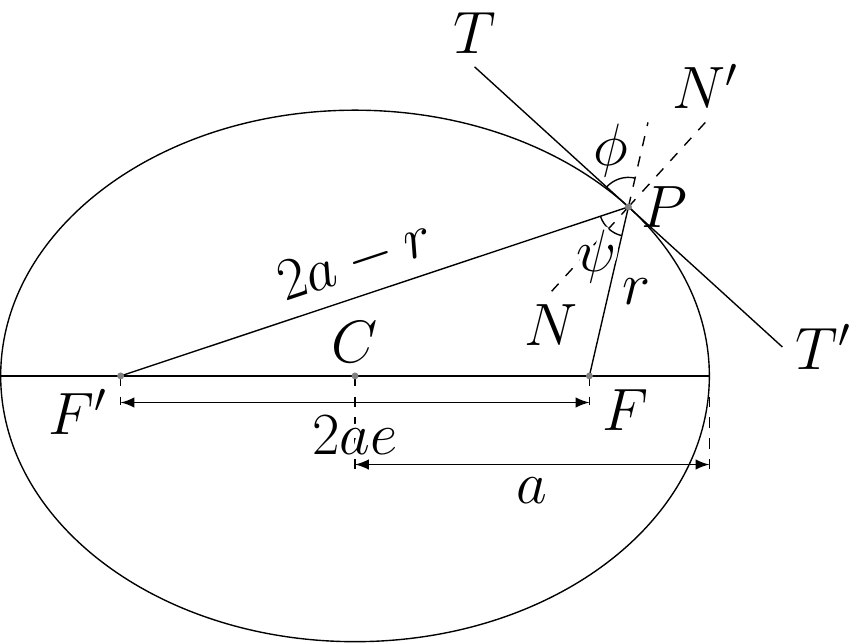}
\caption{Illustration pertinent to the derivation of Eq.~\eqref{eq:PedalEquation} which is an unconventional description of an ellipse, relating $r$ and $\phi$. Point $C$ is the center of the ellipse, and $F'$ and $F$ are the foci. The ellipse has semi-major axis $a$ and eccentricity $e$. The  distance  between  foci  is  $2ae$ and the sum of distances from the foci to any point $P$ on the ellipse is $2a$.}
\label{fig:PedalEquation}
\end{center}
\end{figure}

Consider Fig.~\ref{fig:PedalEquation}, where we have denoted the two foci of the ellipse by $F'$ and $F$. We begin by assuming two important properties of an ellipse.
\begin{enumerate}
\item Given any point $P$ that lies on the ellipse, the sum of the lengths $|PF'|$ and $|PF|$ is constant and equal to $2a$.
\item  If a mirror were made in the shape of an ellipse (with some small extent perpendicular to the plane of the ellipse), a ray of light starting from one of the foci $F$,
  hitting any point on the ellipse $P$ would be reflected to the other focus $F'$, and vice versa. Mathematically, this means that the normal to the ellipse at point $P$ bisects the angle $FPF'$, since the angles of incidence and reflection (as measured from the normal) must be equal.\footnote{These two properties are not independent. The second may be derived from the first, for example, by means of Fermat's principle. However, using both of the properties allows us to simplify the algebra and avoid calculus.}
\end{enumerate}

We are now ready to derive the equation we seek. In Fig.~\ref{fig:PedalEquation}, the tangent to the ellipse at point $P$ is shown as the line $TT'$, and the normal is the line $NN'$. Let us denote the angle between line $FP$ and the tangent $T'T$ by $\phi$. Since the distance to each focus from the center is $ae$, the length of the line $F'F$ is $2ae$. Let us denote the length $|PF|$ by $r$.\footnote{Our re-use of the symbols $\phi$ and $r$ in this section is intentional: we will later see that these indeed correspond to their counterparts from the previous section.} Then by property (1), $|PF'| + |PF| = 2a$, or $|PF'| = 2a - r$. Since $TT'$ and $NN'$ are perpendicular, the angle NPF is $90^\circ - \phi$. Using property (2), we therefore have that $\psi = 2(90^\circ - \phi) = 180^\circ - 2\phi$. By applying the law of cosines to triangle $F'PF$, we obtain
\begin{equation}
  \label{eq:CosineRule}
  (2ae)^2 = (2a - r)^2 + r^2 - 2r(2a - r) \cos(180^\circ - 2\phi).
\end{equation}
Using the trigonometric identities $\cos(180^\circ - \theta) = -\cos\theta$ and $\cos 2\theta = 1 - 2\sin^2 \theta$, and simplifying the result, we obtain
\begin{equation}
  \label{eq:PedalEquation}
  \begin{aligned}
    % (2ae)^2 =& (2a - r)^2 + r^2 + 2r(2a - r) (1 - 2\sin^2 \phi)\\
    % =& (2a)^2 + 4ar + r^2 + r^2 + 2r(2a - r) - 4r(2a - r)\sin^2 \phi\\
    % =& (2a)^2 + (2ar - r^2) \sin^2 \phi\\
    % a^2 (1 - e^2) =& (2ar - r^2) \sin^2 \phi.
    (r^2 - 2ar) \sin^2 \phi = -a^2 (1 - e^2).
  \end{aligned}
\end{equation}
The above equation relating $r$ and $\phi$ describes an ellipse of semi-major axis $a > 0$ and eccentricity $0 < e < 1$.\footnote{Substituting $\sin \phi = p / r$ in Eq.~\eqref{eq:PedalEquation} results in the \emph{pedal equation} of the ellipse,~\cite{noll2002teaching} but we prefer to leave it in this form to facilitate geometric intuition.} It also describes a circle if we substitute $e = 0$, but this is a bit harder to see: noting that the maximum values of the expressions $2ar - r^2$ and  $\sin^2 \phi$ are $a^2$ and 1 respectively, we see that the only way the right hand side can attain a value equal to $-a^2$ is with $r = a$ and $\sin \phi = \pm 1$, which indeed describes a circle.

\section{Kepler's I Law: Elliptical trajectory when $E < 0$}
\label{sec:FirstLaw}

We now make the connection between the physical result of section~\ref{sec:Trajectory} with the mathematical result of section~\ref{sec:PedalEquation}. Let us identify the points $P$ and $F$ of Fig.~\ref{fig:PedalEquation} with the position of the body $m$ at some instant and the location of mass $M$, respectively. We can then identify $r$ and $\phi$ of Eq.~\eqref{eq:Trajectory} with the corresponding symbols in Fig.~\ref{fig:PedalEquation}. If we now compare Eq.~\eqref{eq:Trajectory} and Eq.~\eqref{eq:PedalEquation}, we see that they are indeed the same if
\begin{equation}
  \label{eq:Identifications}
  \begin{aligned}
    a \;=&\; -\frac{GMm}{2E},\\
    \vspace{3pt}\\
    e \;=&\; \sqrt{1 + \frac{2EL^2}{(GM)^2 m^3}}.
  \end{aligned}
\end{equation}
We have therefore shown that when $E < 0$, the trajectory of the body $m$ (``the planet'') is an ellipse with the body $M$ (``the Sun'') at one of the foci, having semi-major axis $a$ and eccentricity $e$ determined from the physical parameters and constants of motion through Eq.~\eqref{eq:Identifications}. The values for $a$ and $e$ agree with results derived in textbooks.~\cite{goldstein2002classical} Given the initial conditions, one may calculate $\vec{L}$ and $E$, and therefore determine the shape and orientation of the ellipse.

If $E > 0$ or $E = 0$, we are unable to match the form of Eq.~\eqref{eq:Trajectory} with Eq.~\eqref{eq:PedalEquation} with the constraints $a > 0$ and $0 \leq e < 1$. Therefore, the trajectory in these cases is not an ellipse. In the cases $E > 0$ and $E = 0$, it is likewise possible to derive equations similar to Eq.~\eqref{eq:PedalEquation} for a hyperbola and a parabola, respectively, and show that we can match Eq.~\eqref{eq:Trajectory} and Eq.~\eqref{eq:PreTrajectory} with them.

\section*{Acknowledgments}
The author wishes to acknowledge amateur astronomers from the Bangalore Astronomical Society for useful discussions and feedback.

\end{document}